\input epsf

\documentclass[twocolumn,floatfix,
showpacs,prd,aps,eqsecnum,tightenlines,draft]{revtex4}
\usepackage{amsmath}

\newcommand {\R}{{\rm R}}

\begin{document}

%\twocolumn[\hsize\textwidth\columnwidth\hsize\csname @twocolumnfalse\endcsname
\preprint{AEI-2001-022, gr-qc/0104063}

\title{The Lazarus project:\\
A pragmatic approach to binary black hole evolutions}

\author{John Baker$^{1,2}$, Manuela Campanelli$^{1,3}$,
and Carlos O. Lousto$^{1,3,4}$}

\affiliation{
$1$ Albert-Einstein-Institut,
Max-Planck-Institut f{\"u}r Gravitationsphysik,
Am M\"uhlenberg 1, D-14476 Golm, Germany\\
$2$ Laboratory for High Energy Astrophysics, NASA Goddard Space Flight Center, Greenbelt, Maryland 20771\\
$3$ Department of Physics and Astronomy, The University of Texas at Brownsville,
Brownsville, Texas 78520\\
$4$ Instituto de Astronom\'{\i}a y F\'{\i}sica del Espacio--CONICET,
Buenos Aires, Argentina}

\date{\today}

%\maketitle

%%%%%%%%%%%%%%%%%%%%%%%%%%%%%%%%%%%%%%%%%%%%%%%%%%%%%%%%%%%%%%%%%%%%%%%%%

\begin{abstract} 
We present a detailed description of techniques developed to combine
3D numerical simulations and, subsequently, a single black hole
close-limit approximation.  This method has made it possible to
compute the first complete waveforms covering the post-orbital
dynamics of a binary black hole system with the numerical simulation
covering the essential non-linear interaction before the close limit
becomes applicable for the late time dynamics.  In order to couple
full numerical and perturbative methods we must address several
questions.  To determine when close-limit perturbation theory is
applicable we apply a combination of invariant {\em a priori}
estimates and {\em a posteriori} consistency checks of the robustness
of our results against exchange of linear and non-linear treatments
near the interface.  Our method begins with a specialized application
of standard numerical techniques adapted to the presently realistic
goal of brief, but accurate simulations.  Once the numerically modeled
binary system reaches a regime that can be treated as perturbations of
the Kerr spacetime, we must approximately relate the numerical
coordinates to the perturbative background coordinates.  We also
perform a rotation of a numerically defined tetrad to asymptotically
reproduce the tetrad required in the perturbative treatment.  We can
then produce numerical Cauchy data for the close-limit evolution in
the form of the Weyl scalar $\psi_4$ and its time derivative
$\partial_t\psi_4$ with both objects being first order coordinate and
tetrad invariant.  The Teukolsky equation in Boyer-Lindquist
coordinates is adopted to further continue the evolution.  To
illustrate the application of these techniques we evolve a single Kerr
hole and compute the spurious radiation as a measure of the error of
the whole procedure.  We also briefly discuss the extension of the
project to make use of improved full numerical evolutions and outline
the approach to a full understanding of astrophysical black hole
binary systems which we can now pursue.

\end{abstract}

%%%%%%%%%%%%%%%%%%%%%%%%%%%%%%%%%%%%%%%%%%%%%%%%%%%%%%%%%%%%%%%%%%%%%%%%%


\maketitle

%\vskip2pc]

%\narrowtext
%\preprint{} 

%%%%%%%%%%%%%%%%%%%%%%%%%%%%%%%%%%%%%%%%%%%%%%%%%%%%%%%%%%%%%%%%%%%%%%%%%

\section{Introduction}

Binary black hole mergers are among the most powerful and
efficient sources of gravitational radiation in our universe
and are thus the primary targets for direct experimental detection 
by the future interferometric observatories. 
Recent astronomical observations of x-ray emission sources
reinforce the evidence of black holes in many galaxies, 
and astrophysical simulations of globular clusters 
\cite{Zwart99,Zwart99b} show binary black holes mergers 
in such an abundance to boost the gravitational wave detection 
rate estimation to $1.6\times 10^{-7}$ $yr^{-1}\,Mps^{-3}$,
which results in about one detection event every 2 years 
for LIGO and in one event per day for LIGO II. 

It is thus not surprising that on the theoretical side the study of
binary black hole mergers has become one of the most exciting and
challenging topics in astrophysical relativity.  Several theoretical
approaches have been developed for treating these systems.  So far the
post-Newtonian approximation (PN), has provided a good understanding
of the early slow adiabatic inspiral, or ``far-limit'', phase of these
systems. Similarly, for the final moments, when black holes are close
enough to each other to sit inside a common gravitational well, one
can successfully apply the ``close limit'' approximation (CL)
\cite{Price94a}, which  effectively describes the whole system as 
a perturbation of a single black hole which rapidly 
``rings-down'' to stationarity.
Before this last stage, though, when the black holes are still close 
to the {\it innermost stable circular orbit} (ISCO), the 
orbital dynamics are expected to yield to
% is expected to change rapidly from inspiral to 
plunge and coalescence. No approximation method can be applied in
this highly nonlinear phase and it is generally expected that one 
can only treat the system by a 
full numerical (FN) integration of Einstein's equations.

Intensive efforts have been underway in the past decade to develop
numerical codes able to solve Einstein's general relativity equations,
by the use of powerful supercomputers.  So far the numerical treatment
of black hole systems in full three dimensions (3D) has proved very
difficult and challenging because of the huge computer memory
requirements, on one hand, and of very severe numerical instabilities,
on the other, which make the codes fail before any useful
gravitational wave information can be extracted.  In spite of such
difficulties, interesting progress has been made, including, for
example the work in \cite{Alcubierre00b}, where a true 3D simulation
based on the traditional $3+1$ decomposition of space and time has
been successfully carried out for the so-called non-axisymmetric
`grazing' collisions of two black holes.  However, because of the
limited evolution time achievable before these codes become unstable
or otherwise inaccurate these simulations must still begin too late in
the plunge to be practical for direct astrophysical application.  In
most cases treatable so far, the close limit approximation theory
represents a good alternative model for the late time dynamics of
these systems.

Considering the above situation, in Refs \cite{Baker00b,Baker:2001nu}
we introduced a new hybrid  approach to the binary black hole merger 
problem, called the {\it Lazarus Project}, with the motivation of providing 
expectant gravitational wave observers with some early estimate of 
the full merger waveforms within a `factor two', and to 
{\it guide} future, more advanced numerical simulations.
The key idea of the Lazarus Project is very simple: 
combine the best of the already existing approaches by applying 
each of these methods in sequence and in their best suited regime, 
while focusing the numerical simulations squarely on the intermediate 
phase of the interaction where no available perturbative approach 
is applicable. 

Clearly, the primary task of the combined model is developing
appropriate interfaces between these three existing treatments in such
a way that we can also benefit of future improvements in any of the
above three approaches.  In an earlier letter \cite{Baker00b} we
presented the first results of our eclectic approach for a model
problem, the head-on collision of black holes, where we successfully
addressed to the problem of combining the close-limit approximation
describing ringing black holes and full three-dimensional numerical
relativity. In this well-known case, our method proved capable of
determining radiation waveforms with accuracy comparable to the best
published 2D numerical results, allowing at the same time a more
direct physical understanding of the collisions and indicating clearly
when non-linear dynamics are important as the final black hole is
formed.  Previous attempts to make a combined use of numerical and
close-limit evolution\cite{Abrahams95b} have been implemented in the
case of two axisymmetric black holes formed by collapsing matter
\cite{Abrahams95d}, using a 2D numerical code and $l=2$ metric
perturbations ({\it \`a la} Zerilli) of the Schwarzschild background
and are not generalizable to full $3D$ simulations.  In
Ref.\cite{Baker:2001nu} we studied the non-axisymmetric coalescence of
equal mass non-spinning binary black holes from the innermost stable
circular orbit (ISCO) down to the final single rotating black hole,
and provided the {\it first}, astrophysically plausible, theoretical
predictions for the gravitational radiated energy, angular momentum,
and waveforms to be expected from these systems.

A sketch of the eclectic approach to the binary black hole
calculation is outlined in the following steps:  (1) First provide a 
description of the early dynamics of the system with an approach, such
as the post-Newtonian method, which is appropriate for slowly moving,
well-separated black hole. A recent interest within the post-Newtonian 
and gravitational wave research community in providing Cauchy data for 
simulations may soon lead to a practical PN-FN interface. 
(2) Extract critical information about the
late-time configuration of this system, and translate this information to
a corresponding solution of the gravitational initial-value problem.
(3) Apply a full 3D numerical simulation of Einstein's equations to 
generate a numerical spacetime covering the non-linear interaction 
region of the spacetime.  The evolution should proceed for long enough 
so that the subsequent evolution of the region exterior to the 
final single remnant black hole can be well approximated by perturbative
dynamics.  (4) At this point we choose a "late-time" slice from the
numerically generated spacetime, extract
$\psi_4=C_{\alpha\beta\gamma\delta}n^\alpha\bar{m}^\beta n^\gamma\bar{m}^\delta$
and $\partial_t\psi_4$,
to quantify the deviation of the numerical spacetime from a Kerr geometry.
Then (5) evolve via the Teukolsky equation,
which governs the dynamics of Kerr perturbations in the time-domain
\cite{Teukolsky73}, long enough to drive all significant radiation into
the radiation zone where it can be interpreted.
Making the greatest possible use of perturbation theory in this way,
not only saves precious three dimensional computational resources, 
concentrating these, for the first time, 
squarely in the intermediate coalescence phase, 
but also provides a new framework to explore and interpret
the interesting new physics that is expected to take place in the 
transition from nonlinear to linear dynamics.

The emphasis of this paper is to realize steps (2)-(5) above and to
describe in detail a general approach to providing the FN-CL
interface.  In Section II we discuss our approach to the full
numerical simulations which we have used to achieve a successful
evolution of truly detached black holes for the first time.  This
discussion naturally divides into two parts a) our preparation of the
initial data, by which we greatly improve the simulations efficiency
and b) our numerical evolution method.

Two important questions arise in implementing the transition, step
(3), from a numerical approach to a perturbative approach. First, how
long must we evolve the system numerically before we can obtain a
reliable description in terms of a single perturbed black hole?  We
use a combination of several independent and complementary indicators
to establish when perturbation theory should begin to work.  In
Section III, we discuss our study of two of such indicators: a) The
speciality invariant, ${\cal S}$, introduced in Ref.\ \cite{Baker00a},
which is exactly equal to 1 for the Kerr geometry with leading
deviations quadratic in the gravitational distortions. b) By
extracting Cauchy data at successive later numerical time slices.
When the system entered the linear regime, the waveforms evolved via
the Teukolsky equation should essentially superpose to each other.
Consequently also a certain level-off of the radiated energy should be
observed.  While a) gives a local measure of the physical distortions
from the Kerr geometry, b) rather depends on the past light cone data.
 
%refer to evolution of Kerr: \cite{Shinkai00a}
The second question is how to identify the single ``background'' black
hole which is emerging in the numerical spacetime.  In order to define
deviations from this background black hole we must be able to relate
it, by an explicit diffeomorphism, to the numerical spacetime. We need
to specify both the spatial coordinates and the time slice, which in
general may be different from the one used to numerically integrate
Einstein equations. This geometrical puzzle is discussed in detail in
Section \ref{coordinates}.  There is in general no geometrically
preferred way to associate the numerical and background spacetimes,
but the first order gauge and tetrad invariance of the perturbative
formalism implies that the results should not depend strongly on small
variations in these choices.

In Section V, we describe how to compute the Cauchy data for the
Teukolsky equation, i.e.  the Weyl scalar $\psi_4$ and its background
time derivative $\partial_t\psi_4$, from the numerical three metric
$g_{ij}$ and extrinsic curvature $K_{ij}$, on the transition Cauchy
hypersurface.  The numerical calculation of the Cauchy data require,
first, a nontrivial identification of an appropriate numerical
`tetrad', which reduces to the (null and complex) tetrad used in the
perturbative calculation in the small perturbation limit.  Second, the
numerical calculation of $\partial_t\psi_4$ is done `on slice' using
Einstein's equations to be consistent with the Boyer-Lindquist time of
the final Kerr black hole.

In Section VI we briefly describe the perturbative Teukolsky equation
and the 2+1 numerical code used to numerically to solve it.  We then
apply all of our techniques, in Section VII, to evolution of a single
Kerr black hole with vanishing shift and maximal slicing to test the
consistency of our method.  Our essentially trivial result is obtained
in a very non-trivial way since our numerical tetrad is not
necessarily aligned with the principal null directions, nor are our
numerical coordinates the Boyer-Lindquist coordinates used in the
perturbative code.  Only after we make the appropriate rotation of the
tetrad and transform the coordinates to reproduce the Boyer-Lindquist
ones we see quadratic convergence to near vanishing outgoing
gravitational radiation.

\begin{figure}
\epsfysize=3.0in \epsfbox{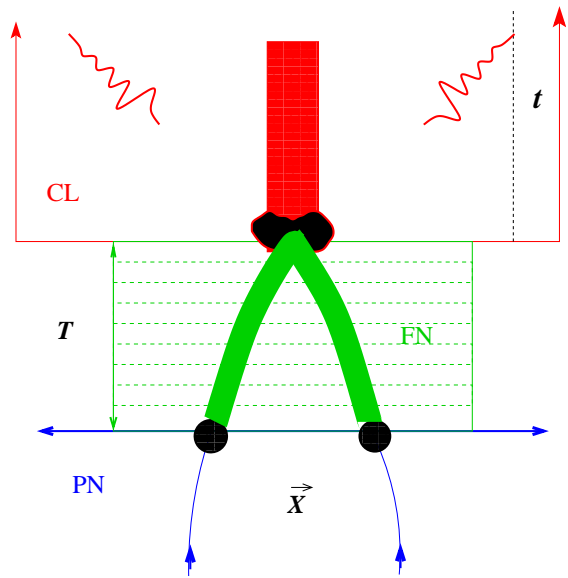}
\caption{The eclectic approach: We represent the three
phases of the binary black hole evolution and the corresponding
techniques adapted to each phase. The full numerical (FN)
evolution is located to cover the truly nonlinear dynamical interaction.
The domain of perturbative evolution (CL) follows
the FN domain allowing indefinite evolution. 
Waveforms are extracted at the dotted world line depicted on the right.
Though such observers are located in the CL part of the spacetime they
will experience all radiation 
arriving from the strong field dynamical FN region. In the far
limit regime we envision to use the post-Newtonian (PN) approximation}
\label{fig:vision}
\end{figure}

%%%%%%%%%%%%%%%%%%%%%%%%%%%%%%%%%%%%%%%%%%%%%%%%%%%%%%%%%%%%%%%%%%%%%%%%%

\section{summary of the full numerical techniques}

In our full numerical simulations we use many of the standard
techniques applied in, for example, the Binary Black Hole Grand  
Challenge effort, with adaptations appropriate the
needs of our more specifically defined numerical simulation problem.
Many previous applications of numerical relativity to
the binary black hole problem have been developmental test problems aiming
toward an ultimate goal of indefinitely long-running 3D numerical
simulations to cover the evolution beginning with well separated black 
holes and evolving through the entire interaction until further radiation
is no longer significant.  With regards to gravitational radiation,
these efforts have been focused on indefinite numerical stability
and successful 
radiation waveform extraction by an observer in the ``far away'' region
of the numerical domain.  These efforts have often been successful with
relatively
brief black hole evolutions, but have demonstrated the serious difficulties
in succeeding with the desired long-running numerical simulations, and 
this approach has not yet generated radiation studies which approach 
relevance to astrophysical problems.

We will ask less of our numerical simulations.  Our demand is for a
highly accurate determination of the most significantly non-linear
part of the binary interaction.  We will try to make use of codes that
may only run stably for a relatively brief period, but which can
provide an accurate representation of the part of the spacetime we are
most interested in.  This point of view allows us, for example, to
avoid the difficult problem of imposing physically accurate outer
boundary conditions, by only considering the part of the spacetime
causally separated from the boundary.  We find that this can be done
much more efficiently in specialized coordinates, described in the
first section below.  Similarly, we have not yet needed more stable
formulations of Einstein's equations, or difficult sophisticated
techniques such as black hole excision. Our straightforward numerical
approach to evolution is described in the second section.

%%%%%%%%%%%%%%%%%%%%%%%%%%%%%%%%%%%%%%%%%%%%%%%%%%%%%%%%%%%%%%%%%%%%%%%%%

\subsection{Preparing the initial data}

Ultimately we wish to derive initial data based on information from
an approximation procedure, such as the post-Newtonian method which is
applicable in the limit of slow-moving/far-apart black holes.  As no
such interface is presently available we use, in our present work, 
initial data from an alternative source, commonly applied in numerical 
relativity, the ``puncture'' formalism with conformally flat three-metric
and purely longitudinal extrinsic curvature on a maximal 
slice. This assumes a three-sheeted topology instead
of an inversion symmetry across the throats\cite{Cook94} allowing for
a solution of the elliptic Hamiltonian constraint equation
without having to impose
interior boundary conditions\cite{Brandt97b}.

Within this family it is possible to identify data roughly
corresponding to quasi-circular orbits using the effective potential
method as in Ref.\ \cite{Cook94}. The binding energy of the system is
computed as a function of the proper separation of the holes keeping
everything else constant. A minimum in the binding energy is then
interpreted as giving an stable quasi-circular orbit. Within this
approach an ISCO is determined by varying the orbital angular momentum
of the system until this minimum becomes an inflection point.  For
less separated configurations, a stable quasi-circular orbit is no
longer possible.  We use these ISCO data, determined in
\cite{Baumgarte00a}, for non-spinning equal-mass black holes as a
particularly reasonable starting point for approaching astrophysical
systems\cite{Baker:2001nu}.

Having selected the physical initial data we then prepare it for
numerical evolution.  When Smarr and York \cite{Smarr78b} spelled out
the problem of 3+1 numerical relativity in the 1970's, they
specifically sought out methods which would be invariant to gauge
transformations in the initial data.  In the pursuit of long-running
all-purpose numerical relativity tools, this viewpoint has been
traditionally preserved, and little attention has been given to the
question of choosing appropriate coordinates for the initial data.  It
is clear though, that whenever differential equations are to be solved
numerically, some choices of variables (coordinates) will be more
practical than others.  In a wave-propagation problem, for instance,
the simulation will be much more efficient if a wave is evenly
resolved as it moves across the numerical domain, or similarly, if the
coordinate characteristic speeds were constant in space and time.

\begin{figure}
\epsfysize=2.7in \epsfbox{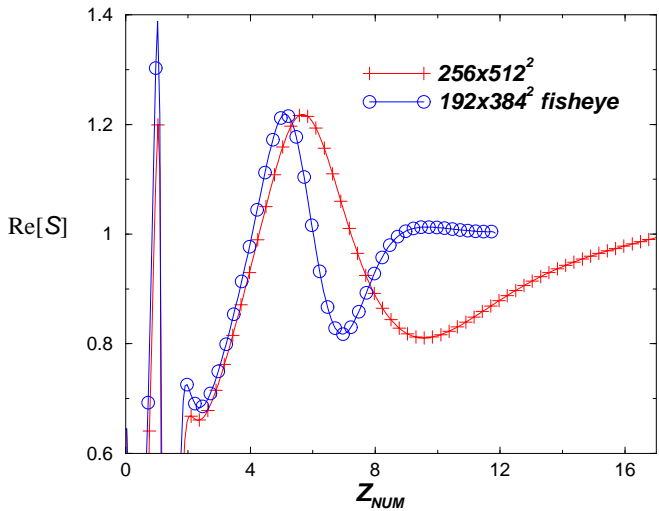}
\caption{The benefit of our fish-eye coordinates compared against the 
typical isotropic coordinates.  The $\cal{S}$ invariant, plotted here,
gives an indication of the radiation moving out from an initial ISCO
system after $10M$ of numerical evolution. In the strong field region
up to $z=6$ the two coordinate systems are very similar.  Moving
outside that region though the Fish-eye coordinate cover a
significantly larger region of the physical spacetime with fewer
grid-points.  The extra grid-points in isotropic coordinates are
wasted by over-resolving the outer part of the radiation.  In Fish-eye
coordinates the wave is resolved more evenly.}
\label{fig:FE}
\end{figure}

For numerical relativity simulations in practice we are often very far
from this ideal.  In typical coordinates, such as isotropic
coordinates for our (initially) conformally flat spaces, the waves are
strongly red-shifted as they move away from the strong-field region.
Since we require both a physically large computational domain and also
high resolution in the strong field region, use of the standard
coordinates leads to a great waste of numerical effort on
over-resolving an outgoing radiation wave which was originally
generated with much poorer resolution.  In this way, relatively little
is gained by, expanding the computational domain with additional
numerical grid-points.  We find that we can make great improvements in
numerical efficiency with a relatively simple {\it ad hoc} coordinate
transformation on the initial data which we call `fish-eye'
coordinates.  A typical such transformation is a radial rescaling,
$r_{iso}=R_{num}\cosh{((R_{num}/R_0)^n)}$ with typical values
$R_0=7.7$ and $n=2$.  This allows us to maintain a central resolution
of up to $M/24$ with outer boundaries near $r_{new}=37M$ using only
$256\times512^2$ grid-points, moving the outer boundary much farther
away without loss of physical resolution in the strong field region.
This problem is illustrated in Figure \ref{fig:FE}, which shows data
from numerical simulations in two alternative coordinate systems after
$10M$ of evolution from an initial ISCO configuration.  The outgoing
radiation wave is noticeable in the real part of the gauge-independent
$\cal{S}$ invariant discussed in Sec. \ref{Sinvsec}.  These curves
represent the same physical spacetime as seen from alternative
numerical coordinate systems.  In this figure the strong field
dynamics are most important in the left side on the figure up to about
the value of the numerical coordinate (along the z-axis) of
$Z_{NUM}=6$.  Up to that point the two coordinate systems are nearly
identical.  As we add grid-points on the right side of the figure
beyond this strong field region, we are frustrated, in the isotropic
coordinate case by the red-shift effect and only modest additional
part of the outgoing wave, about half a wave cycle, is added to the
grid when we roughly triple the grid dimension.  In the case of our
fish-eye coordinate the wave is evidently more evenly resolved across
the grid, and we cover the domain of the isotropic coordinate system
with only about a 60\% increase in the grid dimension.  As shown, the
fish-eye coordinate system has a much more distant {\em physical}
outer boundary than that of the isotropic case while still having only
about half of the grid-points in 3D.  Note that we would gain no
further advantage by attempting to compactify spatial infinity as in
for example
\cite{Garfinkle:2000hd} since resolution must
nevertheless still fail to resolve waves
at a finite radius in such a scheme.

Foreseeing longer term full numerical evolutions we have also implemented
other re-coordinatizations of the initial data that have a fairly constant
high resolution in the center of the grid
(where the grid stretching
is more severe) and a lower resolution near the boundaries, but
still fairly constant to allow the application of the usual radiative
boundary conditions (adapted to the different characteristic speed).
One of such functions is
\begin{eqnarray}
r_{iso}&=&R_{num}\left(1+b\left(\tanh\left(\frac{2(R_{num}-R_0)}{d}+0.35
\right)\right.\right.\nonumber\\
&&\left.\left.+\tanh\left(\frac{2R_0}{d}-0.35\right)\right)\right)^2
\end{eqnarray}
with $b,d,R_0$ adjustable parameters that determine the ratio of central
to boundary resolutions, the width and location of the effective resolution
transition region respectively.

%%%%%%%%%%%%%%%%%%%%%%%%%%%%%%%%%%%%%%%%%%%%%%%%%%%%%%%%%%%%%%%%%%%%%%%%%

\subsection{Numerical evolution}

Our numerical evolution must be consistent with our need for highly
accurate relatively brief simulations.  Consequently, in our work so
far, we have used the standard ADM (Arnowitt-Deser-Misner) formulation
of Einstein equations \cite{Arnowitt62} as adapted by Smarr and York
\cite{Smarr78b}. Our evolution equations are thus simply:
\begin{eqnarray}
\hat\partial_0g_{ab}&=&-2\alpha K_{ab}\\
\hat\partial_0K_{ab}&=&-\nabla_a\nabla_b\alpha\nonumber\\
&&+\alpha(R_{ab}-2K_{ac}K^{c}_{b}+K_{ab}K)
\end{eqnarray}
where $\hat\partial_0=\partial_t-\pounds_\beta$. Here and below Latin
indices run from 1 to 3.

Though a newer conformal formulation of Einstein's equations has been
found to be more stable in various numerical simulations
\cite{Alcubierre99d}, here we focus on the accuracy
% and convergence
of the solutions rather than long term stability.
Our observation is that the standard ADM equations seem to 
give more accurate results for binary black hole simulations 
in our gauge while the simulation is stable.

If it is possible to have a slicing which is consistent with that of
our perturbation theory, then we can avoid a rather large technical
problem of producing data on a slice inconsistent with the background.
Consistent with our choice of Boyer-Lindquist coordinates in our
perturbation treatment of the background black hole, we have chosen
maximal slicing to define the lapse $\alpha$,
\begin{equation}\label{Maximal}
K=0, ~~~~\Delta\alpha = \alpha \ K_{ab}\ K^{ab}.
\end{equation}
This implies an elliptic equation for $\alpha$ which we typically have
solved every 5 time steps using Dirichlet boundary conditions.  For
simplicity we set the shift $\beta^i=0$, which is an adequate
condition for relatively brief runs.  The numerical evolution is
performed using an iterative Crank-Nicholson method of third order
which is second order convergent.  In our simulations we have used a
resolutions up to $dx=M/24$ with $dt=0.25\ dx$.  Because we have moved
the outer boundary to a point causally separated from the region we
are interested in it is acceptable simply to impose static boundary
conditions.

In evaluating the results of our numerical simulations we make
frequent use of two indicators: The degree of satisfaction of the ADM
constraint equations gives a measure of the numerical error produced
by the evolution
\begin{eqnarray}
&&\nabla^a(K_{ab}-g_{ab}\ K)=0\\
&&R-2K_{ab}\ K^{ab}+K^2=0.
\end{eqnarray}
These quantities provide an important indication of when numerical
inaccuracies (and eventually instabilities) have become significant in
our simulations.  Even if Einstein's equations could be solved
perfectly, any simulation with a finite boundary is subject to an
additional type of error arising from inappropriate boundary
conditions.  A geometrically correct solution may have physically
unreasonable disturbances propagating in from the boundary. We have
found the speciality invariant $\cal{S}$ \cite{Baker00a} to be a
sensitive indicator such boundary waves, which do not violate the
constraints.

%%%%%%%%%%%%%%%%%%%%%%%%%%%%%%%%%%%%%%%%%%%%%%%%%%%%%%%%%%%%%%%%%%%%%%%%%

\section{Determining the linear regime} 

Black hole perturbation theory has recently generated much interest as
a model for the late stages of a binary black hole collision spacetime
\cite{Price94a}.  When two black holes are close enough to each other
one can simply treat the problem, in the `close limit' approximation,
as a single distorted black hole that `rings down' into its final
equilibrium state.  So after some nonlinear numerical evolution of the
full Einstein's equations for a system of two initially well-detached
black holes, there should always be a transition time, $T$, after
which the system simply behaves linearly i.e. satisfy the linear
perturbation equations around the final Kerr black hole.  Finding the
{\it linearization time} $T$ is thus the first nontrivial question
which arises in the context of our `eclectic' approach.  In other
words, we need one or more working criteria for {\it when} we can
expect perturbation theory to be accurately effective based only on
numerical data. As we shall see below, we apply at least two
independent criteria for estimating the onset of linear dynamics, the
speciality invariant prediction based only on the Cauchy data and
another estimate based on the stability of the radiation waveform
phase.

%%%%%%%%%%%%%%%%%%%%%%%%%%%%%%%%%%%%%%%%%%%%%%%%%%%%%%%%%%%%%%%%%%%%%%%%%

\subsection{The speciality invariant test}\label{Sinvsec}

Motivated by this purpose in Ref.\ \cite{Baker00a} we introduced an
invariant quantity,
\begin{equation}
{\cal S}=27{\cal J}^2/{\cal I}^3,
\end{equation}
where ${\cal I}$ and ${\cal J}$ are the two complex curvature
invariants ${\cal I}$ and ${\cal J}$, which are essentially the square
and cube the self-dual part, $\tilde C_{\alpha\beta\gamma\delta}=
C_{\alpha\beta\gamma\delta}+(i/2)\epsilon_{abmn}C^{mn}_{\,\,\,\,cd}$,
of the Weyl tensor:
\begin{equation}
{\cal I}=\tilde C_{\alpha\beta\gamma\delta}
\tilde C^{\alpha\beta\gamma\delta} ~~{\rm and}~~
{\cal J}=\tilde C_{\alpha\beta\gamma\delta}
\tilde C^{\gamma\delta}_{\,\,\,\,\mu\nu}\tilde C^{\mu\nu\alpha\beta}.
\end{equation}
Both these scalars can be expressed in terms of the Weyl components,
for an arbitrary tetrad choice:
\begin{eqnarray}
{\cal I}&=&3{\psi_2 }^2-4\psi_1\psi_3+\psi_4\psi_0, \nonumber \\
{\cal J}&=&-\psi_2^3+\psi_0\psi_4\psi_2 + 2\psi_1\psi_3\psi_2-\psi_4\psi_1^2
    -\psi_0\psi_3^2.
\end{eqnarray}

The geometrical significance of ${\cal S}$ is that it measures the
deviations from algebraic speciality (in the Petrov classification of
the Weyl tensor).

For the unperturbed algebraically special (Petrov type D) Kerr
background ${\cal S}=1$.  However, for interesting spacetimes
involving nontrivial dynamics, like distorted black holes, which are
in general not algebraically special (Petrov type I), we expect ${\cal
S}=1+\Delta{\cal S}$, and the size of the deviation $\Delta{\cal
S}\neq 0$ can be used as a guide to predict the applicability of black
hole perturbation theory.  In particular we adopt the criterion that,
when ${\cal S}$ differs from its background value of unity by less
than ``a factor of two'' outside the (background) horizon, a
perturbative treatment may be expected to provide a reasonable
description of the radiative dynamics.  A larger deviation from
algebraic speciality implies significant ``second order''
perturbations.  In fact, for perturbations on a background Kerr
spacetime, with an arbitrary tetrad perturbation, one can easily
deduce
\begin{equation}
{\cal S}=1-3\epsilon^2\frac{\psi_0^{(1)}\psi_4^{(1)}}
{(\psi_2^{(0)})^2}+{\cal O}(\epsilon^3), \label{S2}
\end{equation}
where $\psi_0$, $\psi_4$ and $\psi_2$ are the usual Newman-Penrose
complex Weyl scalars.  The lowest order term in the deviation is
second order in the perturbation parameter $\epsilon$, and should tend
to vanish if first order perturbation theory is appropriate.  Note
that the superscripts ${(0)}$ and ${(1)}$ stand respectively for
background and first order pieces of a perturbed quantity, where
$\epsilon$ is a perturbation parameter.

\begin{figure}
\epsfysize=3.0in \epsfbox{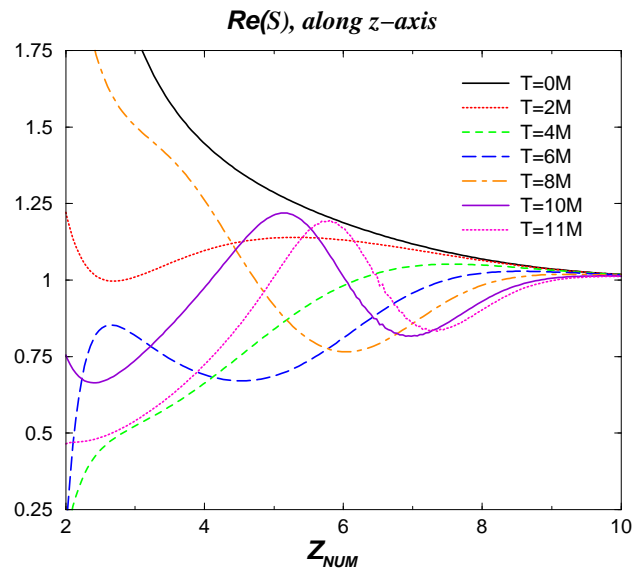}
\caption{The `speciality' invariant for binary black holes evolving from
the `ISCO' showing damped oscillations around unity, its Kerr value.
The location of the horizon in these coordinates is roughly $2.5$.
Its behavior at larger radius suggests radiation is beginning to leave
the system.}

\label{fig:rS}
\end{figure}

In Fig.\ \ref{fig:rS} we display the speciality invariant along the
z-axis, perpendicular to the orbital plane of two black holes starting
the evolution from the ISCO determination used in\cite{Baker:2001nu}.
Its value oscillates around one (the Kerr background value).  After
some evolution $T\approx11M$, the amplitude of the oscillation
decreases to a deviation below 50\% outside the horizon (located at
around $Z_{NUM}\approx2.5$ in the numerical coordinates, and
perturbation theory can reliably take over the remaining of the
evolution. Because the gravitational field has two degrees of freedom
is it clear that the $\cal{S}$-invariant alone is insufficient to
provide a complete description of black hole perturbations, and can be
complemented with its time variation $\dot{\cal S}$.  Consequently, we
have been looking at the turning points where $\dot{\cal S}=0$ and the
amplitude of the distortion reaches a maximum.

As noted in Section III, the ${\cal S}$ is also very useful outside
the perturbative context.  Its usefulness is derived from the fact
that it is a gauge invariant quantity which, unlike $I$ and $J$ is not
dominated by strong ``peeling property'' fall-off behavior, which
tends to indicate spacetime dynamics.  Because the Weyl tensor,
$C_{\alpha\beta\gamma\delta}$, carries information about the
gravitational fields in the spacetime, ${\cal S}$ turns out to be an
interesting indicator of radiation of the spacetime and tests, for
instance, how much radiation is produced by the imposition of
approximate boundary conditions. We have found that the ${\cal
S}-$invariant is simple to calculate and can be applied directly to
full 3D numerical evolutions to provide a gauge invariant indication
of the dynamics.

%%%%%%%%%%%%%%%%%%%%%%%%%%%%%%%%%%%%%%%%%%%%%%%%%%%%%%%%%%%%%%%%%%%%%%%%%

\subsection{Waveform locking and energy plateau}

\begin{figure}
\epsfysize=2.3in \epsfbox{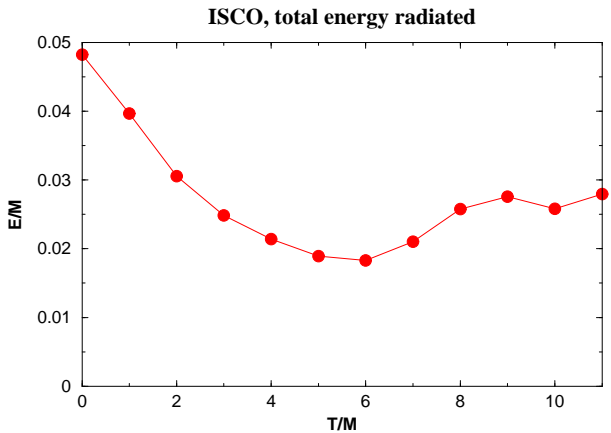}
\caption{Energy radiated from two black holes from ISCO configuration
for different transition times showing a plateau when reaching the
`linear' regime.}
\label{fig:plateau}
\end{figure}

The phase and the amplitude of the radiation, or equivalently the
locking of the waveforms and the corresponding energy plateau, also
provide an indicator of linear dynamics.  Starting with detached black
holes, we expect an initial period of weak bremsstrahlung radiation
followed by the appearance of quasi-normal ringing. On the other hand,
switching to perturbative evolution immediately leads to premature
ringing. Hence if we cut short the numerical simulation and apply
linear theory too early, we observe quasi-normal ringing too early and
calculate a waveform which is out of phase with the desired result.
Comparing waveforms derived from differing durations of numerical
simulation then we tend to see a phase shift in the onset of the
ringing when we have not yet allowed enough numerical simulation.  In
practice, we thus follow the behavior of the waveforms through the
evolution by extracting the Cauchy data at successive later numerical
time slices. When the system enters in the linear regime, the
waveforms evolved via the perturbative Teukolsky equation should
essentially superpose to each other, as changing this transition time
amounts to an equivalent exchange of linear and non-linear evolution
for the intervening region of spacetime.  Consequently also a certain
level-off of the radiated energy should be observed (See Fig.\
\ref{fig:plateau}).

\begin{figure}
\epsfysize=3.0in \epsfbox{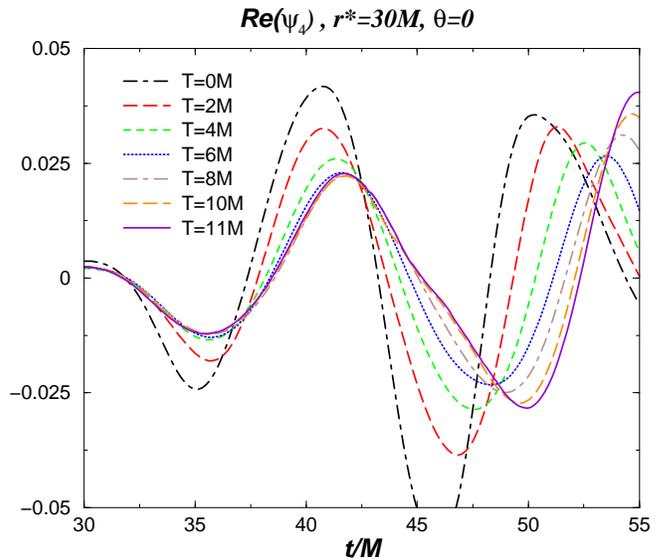}
\caption{Detail of the progressive waveform locking process for
black holes at the location of the ISCO.} 
\label{fig:wfP3}
\end{figure}

As we show in Fig.\ \ref{fig:wfP3}, extracting waveforms every $1M$ of
non-linear numerical evolution allows us to study the transition to
linear dynamics, and to perform important consistency tests on our
results.  If we have made a good definition of the perturbative
background, as described in Section \ref{coordinates}, then we can
expect our radiation waveform results to be independent of the
transition time, $T$, once the linear regime is reached and for as
long as the numerical simulation continues to be accurate.

A closer look at Fig.\ \ref{fig:wfP3} gives us an idea of how the
linearization happens. Curves of $T=10\&11M$ of evolution are close to
the correct waveform for this orbital case starting at a proper
separation $L/M=4.9$. If we apply right away the close limit
approximation we get the curve labeled by $T=0M$ which starts ringing
prematurely. After $2M$ of full numerical evolution we obtain good
agreement with the correct waveform up to $t/M\approx33$. When
perturbation theory takes over after $4M$ of full numerical evolution
the agreement is very good up to $t/M\approx38$. Near $t/M=45M$ we
seen we need $8M$ of nonlinear evolution while near $t/M=50M$ $10M$ of
full numerical evolution are needed and at longer times the agreements
begins to be fine for the whole relevant waveform.  This process shows
how the full nonlinear dynamics shifts to a central region covered by
the common potential barrier allowing to describe linearly the
evolution of the outer part.

%%%%%%%%%%%%%%%%%%%%%%%%%%%%%%%%%%%%%%%%%%%%%%%%%%%%%%%%%%%%%%%%%%%%%%%%%

\subsection{Common horizon}

An intuitive picture to visualize the applicability of the close limit
approximation would be the appearance of a common event horizon that
encompasses the binary system. Under these conditions the spacetime
exterior to the horizon (the relevant one for computing gravitational
radiation reaching infinity) can be treated as perturbations of a Kerr
hole. In practice {\it event} horizons are difficult to compute in
numerical relativity because they are a global feature of the
spacetime and we would need to first evolve the binary system for a
long time and then extract a {\it posteriori} the information to
locate the event horizon. An easier quantity to compute is the {\it
apparent} horizon that can be defined locally as the outermost
marginally trapped surface of the spacetime where a congruence of null
rays directed outwards have vanishing expansion
\cite{Alcubierre98b}. A common apparent horizon lies inside and in a
binary system appears later than a common event horizon; and typically
much later than when the system can be effectively described by linear
perturbations.  The linearization time refers to when the close limit
approximation can be applied and this intuitively occurs when a common
potential barrier covers the binary system. In black hole perturbation
theory, a potential is present somewhat outside the horizon of the
black hole which tends to prevent radiation from escaping this region.
This is the main reason why the close limit is such a good
approximation even beyond original expectations \cite{Gleiser96b}.

%%%%%%%%%%%%%%%%%%%%%%%%%%%%%%%%%%%%%%%%%%%%%%%%%%%%%%%%%%%%%%%%%%%%%%%%%

\section{Constructing the Kerr background}\label{coordinates}

Einstein's theory of gravity in principle demands the equivalence of
all coordinate representations of gravitational dynamics. However, in
practice one always needs to choose a convenient gauge to accurately
carry over the full numerical evolution to the point where the two
black hole system effectively behaves like a single perturbed black
hole.  Having determined that a late time numerical spacetime geometry
is close to the Kerr spacetime does not give us any information about
the coordinate system in which this is written.  In order to be able
to continue the numerical evolution with the Teukolsky equation\
(\ref{Teuk}), we thus need to reconstruct a Kerr background in a
recognizable form, for instance in Boyer-Lindquist
coordinates. Because there is in general no unique procedure to
reconstruct such a Kerr background, we shall require that this should
be {\it close} enough to the given numerical spacetime.  In other
words, we will require that the two spacetimes agree {\it to the first
order} in $\epsilon$.  Since the physics of our problem will then be
described by quantities, like $\psi_4$, which are first order gauge
(and tetrad) invariant, the physical results we compute will be
independent (to first order) of the identification of the background
coordinates we describe below.  To have complete theoretical control
of the perturbation theory, it is desirable to have to a complete
family of initial data sets which reduces to the background geometry
in the limit $\epsilon\rightarrow0$.  While this requirement is not
strictly required in a practical perturbative application
\cite{Abrahams95b}, we would like to stay as close as possible to this
arrangement for its benefit in evaluating our results.  In our case
the perturbation parameter $\epsilon$ can be regarded as a decreasing
function of the transition time $T$.  In practice, we will not be able
to achieve an exact Kerr black hole in the $T\rightarrow\infty$ limit,
but we will aim for the practical goal that the remaining
perturbations are small compared to the radiation we are interested
in, a condition which we test in Section\ \ref{Kerr}.

We initially suppose that the background Kerr black hole is given by the
parameters $M$ and $a$ of the initial data. With a first estimate of
the total radiated energy and angular momentum these parameters can be
iterated to approach the final values for the stationary Kerr black hole.

The Kerr metric in Boyer--Lindquist coordinates $(t,r,\theta,\phi)$
reads,
\begin{eqnarray}
&&
ds^2=-\left(1-\frac{2Mr}{\Sigma}\right) dt^2 
+\left(\frac{\Sigma}{\Delta}\right) dr^2\nonumber\\
&&
+ \Sigma d\theta^2+\sin^2\theta\frac{\Omega}{\Sigma} d\phi^2
-\frac{4aMr}{\Sigma}\sin^2\theta dtd\phi,
\end{eqnarray}
where
$\Delta=r^2-2Mr+a^2$,$\Sigma=r^2+a^2\cos^2\theta$ and
%$\Omega=(r^2+a^2)^2-\Delta a^2\sin^2\theta$
$\Omega=(r^2+a^2)\Sigma+2M r a^2\sin^2\theta$, $M$ is the mass 
of the black hole, $a$ its angular momentum per unit mass.

%%%%%%%%%%%%%%%%%%%%%%%%%%%%%%%%%%%%%%%%%%%%%%%%%%%%%%%%%%%%%%%%%%%%%%%%%

\subsection{The slice}

\begin{figure}
\epsfysize=3.0in \epsfbox{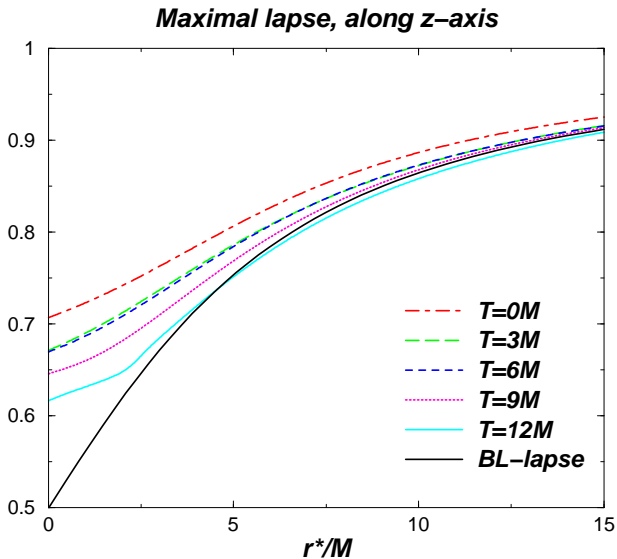}
\caption{The maximal lapse used for black holes evolving
from `ISCO' compared to the analytic Kerr lapse in Boyer-Lindquist
coordinates for different evolution times.}
\label{fig:lapse}
\end{figure}

We recall that a Boyer-Lindquist slice of the Kerr metric has $K=0$.
The full numerical coordinate condition of Maximal slicing, Eq.\
(\ref{Maximal}), is solved for the lapse $\alpha$ with an exterior
boundary condition set to reproduce the value of the Boyer-Lindquist
lapse there, but to vanish at the location of the individual black
hole ``punctures''. The resulting lapse from the evolution of two
holes from `ISCO' is shown in Fig.\ \ref{fig:lapse}. The lapse
resembles the Boyer-Lindquist lapse initially and further evolution
bring them closer.  Thus the maximal lapse with our boundary condition
approaches the background lapse quite closely.  Where there are
differences, near the horizon, our lapse tends to produce a coordinate
system in which the coordinate observers drift slowly into the black
hole.  Considering our coordinate trajectories from the frame of the
background black hole, one can conclude that since the trajectories
and lapse are similar away from the horizon, and the lapse is a bit
different near the horizon, our slicing will be close to the
background slicing, but slightly distorted toward the future near the
horizon.  In Section VI we try to quantify the significance of this
distortion with a numerical study of the Kerr spacetime in these
coordinates.

Other lapse possibilities can be considered which also produce a
slicing similar to that of the Boyer-Lindquist background, algebraic
slicings, for instance \cite{Anninos94c} `$(1+log)$', and a
re-parametrization of the maximal slicing by an $f(\alpha)$ such that
the numerical lapse resembles even closer the Boyer-Lindquist one.  We
performed such tests and check that whenever the deviations from the
Boyer-Lindquist lapse are close enough the results for radiated
waveforms and energies do not change notably, in agreement with the
first order gauge invariance of $\psi_4$ and $\partial_t\psi_4$.

%%%%%%%%%%%%%%%%%%%%%%%%%%%%%%%%%%%%%%%%%%%%%%%%%%%%%%%%%%%%%%%%%%%%%%%%%

\subsection{The spatial coordinates}
\label{ssec:spatial_coordinates}

The general idea here is to numerically compute physical quantities or
geometrical invariants and relate them to their analytic expressions
in the perturbatively preferred coordinate system. Curvature invariant
methods have the distinct advantage that they can be applied to
evolutions using numerically generated coordinates which are not
understood analytically.  On the other hand, the values of curvature
invariants in the perturbed spacetime may be sensitive to perturbative
distortions, making them less useful for identifying a background
spacetime.  In light of these effects, we pursue a combined approach,
utilizing both gauge and geometrical information where each seems most
appropriate.  In the outer regions of our spatial slices we expect the
gauge to be close to the quasi-isotropic gauge for Kerr data.  Moving
in from this to the interior region we expect, most importantly, two
gauge effects.  First, our slicing has the tendency (without a shift)
to cause the coordinates to fall inward with evolution.  We counteract
this with a rescaling of the radius $r_{\rm Kerr}=r_{\rm Kerr}(r)$,
making use of the ${\cal I}$ invariant which depends most
significantly on the radial coordinate in the background slice, ${\cal
I}=3M^2/(r-ia\cos\theta)^6$.  We use this relation and information
about the numerical value of ${\cal I}$ to define the rescaled
radius. To do this we need to produce one value of ``${\cal I}$'' for
each constant $r$ sphere in the numerical slice. The maximum value of
$\int_0^{2\pi}{\cal I}d\varphi$ tends to lie on the equatorial
symmetry plane of our binary black hole problem, where the
$\cal{S}$-invariant also indicates relatively weaker distortions.
This makes
\begin{eqnarray}
<{\cal I}> &=& \frac{1}{2\pi}\int_0^{2\pi}
{\cal I}(r,\theta=\pi/2,\varphi)d\varphi\\
r_{\rm Kerr}&=&\sqrt[6]{3M/<{\cal I}>}
\end{eqnarray}
a practical definition which counteracts the coordinate infall.

Unlike for $r$ there are no obvious dynamical effects on the $\theta$
coordinate and it has been sufficient to adopt the numerical value
$\cos{\theta} = z/\sqrt{(x^2 + y^2 + z^2)}$.  We successfully applied
these re-mapping of coordinates already in the head-on collision
case\cite{Baker00b}.

The second important coordinate effect, which becomes relevant when
the total angular momentum is significant, is the result of frame
dragging caused essentially by the difference between our vanishing
shift, and the non-vanishing Boyer-Lindquist shift.  This effect drags
the coordinates in the $\varphi$ direction and has the effect of
producing an off-diagonal distortion in the numerical metric.  We can
undo the frame dragging by attempting to restore the diagonal form of
the Boyer-Lindquist three metric.

We seek to set frame-dragging gauge freedom by supplementing the
Cartesian definition of $\varphi \equiv \arctan[y/x]$ with a
correction that makes the metric component most strongly affected by
frame dragging, $g_{r\phi}$, vanish,
\begin{equation}\label{phi-transformation}
\phi=\arctan[y/x]+\int (\hat{g}_{r\varphi}/\hat{g}_{\varphi\varphi}) dr,
\end{equation}
where the `hat' stands for the full numerically evolved metric.  To
see this, consider $\hat{g}_{ab}$ with no shift, and transform to
$g_{ab}$ with $\phi$-shift by $\varphi\to\phi=\varphi+\varphi_{\rm
offset}(t,r,\theta)$ (note that $g_{\phi\phi} =
\hat{g}_{\varphi\varphi}$). Then, since the Kerr three-metric is
diagonal,
\begin{equation}
g_{r\phi}=0=\hat{g}_{r\varphi}
-\partial_r\varphi_{\rm offset}\ \hat{g}_{\varphi\varphi}
\end{equation}
so that
\begin{equation}
\partial_r\varphi_{\rm offset}=\frac{\hat{g}_{r\varphi}}
{\hat{g}_{\varphi\varphi}}
\end{equation}

Similarly, since the numerical metric has zero shift, we find
\begin{equation}
g_{t\phi}=\hat{g}_{t\varphi} -\partial_t\varphi_{\rm offset}\
\hat{g}_{\varphi\varphi} = -\partial_t\varphi_{\rm offset}\
\hat{g}_{\varphi\varphi}
\end{equation}
so that
\begin{equation}
\partial_t\varphi_{\rm offset}=-\frac{g_{t\phi}}
{\hat{g}_{\varphi\varphi}}=-\frac{N^{\phi}_{\rm Kerr}\ g_{\phi\phi}}
{\hat{g}_{\varphi\varphi}}=-N^{\phi}_{\rm Kerr}.
\end{equation}
Since $N^{\phi}_{\rm Kerr}$ is constant in $t$,
\begin{equation}
\partial^2_{t}\varphi_{\rm offset}=0,
\end{equation}
\begin{equation}
\varphi_{\rm offset}= -t N^{\phi}_{\rm Kerr}.
\label{offset}
\end{equation}

\begin{figure} 
\epsfysize=3.0in \epsfbox{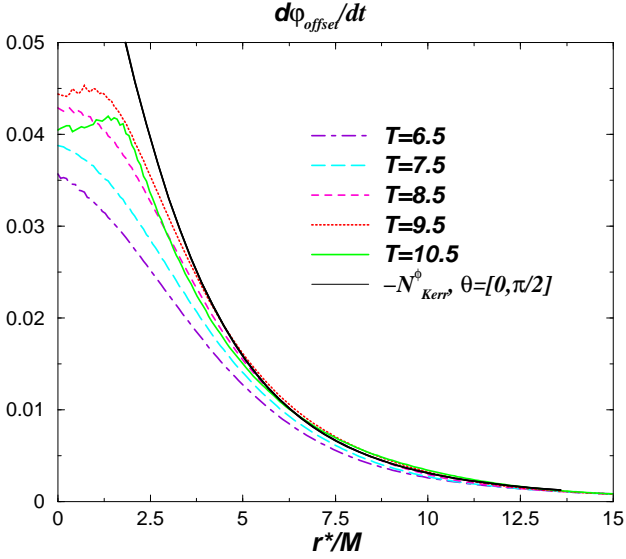}
\caption{The effective shift correction for black holes evolving
from `ISCO' compared to the analytic Kerr shift in Boyer-Lindquist
coordinates for successive evolution times.}
\label{fig:shift}
\end{figure}

Equation (\ref{offset}) allows us to test how close our derived (from
the block diagonal metric condition) shift correction is to the
Boyer-Lindquist shift. The results of this comparison are displayed in
Fig.\ \ref{fig:shift}. For two black holes evolving from the ISCO, the
shift correction correctly reproduces the frame dragging effect
outside the potential barrier of the system and evolution bring the
shift closer to that of a single rotating Kerr hole.

We note that some means of fixing this frame-dragging degree of gauge
freedom, as we have done here, is essential also if one wishes to
speak meaningfully of the number of orbits the system has undergone in
the strong field region during numerical simulations.

As already pointed out there is no unique way of choosing the
coordinate transformations in order to bring them closer to that of
the Kerr background. Our philosophy in this section has been to
consider the simplest of these transformations that approaches the
Boyer-Lindquist coordinates with enough accuracy for the binary black
hole numerical simulations we are interested in. Obviously, other
possibilities that would improve the accuracy of the procedure can be
incorporated as needed.  We also note that the optimal choice of
coordinate transformation needed here may depend on the shift
condition used in evolution and the coordinates used for the initial
data.  The use of a shift condition, such as minimal distortion (with
the appropriate boundary conditions), which is naturally adapted to
the stationarity Killing vector of the background Kerr spacetime
\cite{Smarr78b} may, for example, eliminate frame dragging and 
thus reduce the need for a transformation such as (\ref{phi-transformation}).

%%%%%%%%%%%%%%%%%%%%%%%%%%%%%%%%%%%%%%%%%%%%%%%%%%%%%%%%%%%%%%%%%%%%%%%%%

\section{Constructing the Cauchy data}

Given the numerical metric $g_{ij}$ and the extrinsic curvature
$K_{ij}$ derived as in Section II on a Cauchy hypersurface, and the
coordinates of the background metric determined in Section IV, we
proceed to compute the Weyl scalar $\psi_4$ and its background time
derivative $\partial_t\psi_4$, the Cauchy data we will need to
continue the evolution via the Teukolsky equation.  As was discussed
in references
\cite{Campanelli98a,Campanelli98b,Campanelli98c,Campanelli99}, 
one can make the following 3+1 decomposition, using the basis
$\theta^0=dt,~\theta^i=dx^i+N^idt$, to get
\begin{eqnarray}
&&\psi _4 = \left[ {R}_{ijkl}+2K_{i[k}K_{l]j}\right]
{n}^i\bar{m}^j{n}^k\bar{m}^l \nonumber \\
&&-8\left[ K_{j[k,l]}+{\Gamma }_{j[k}^pK_{l]p}\right]
{n}^{[0}\bar{m}^{j]}{n}^k\bar{m}^l \nonumber \\
&&+4\left[ {R}_{jl}-K_{jp}K_l^p+KK_{jl}\right]
{n}^{[0}\bar{m}^{j]}{n}^{[0}\bar{m}^{l]}, \label{psi4}
\end{eqnarray}
and its time derivative
\begin{eqnarray}
&&\partial _t\psi _4 = -N^i\partial_i\left(\psi_4\right) 
+\left[ {\hat\partial }_0R_{ijkl}\right]
{n}^i\bar{m}^j{n}^k\bar{m}^l 
\nonumber \\
&&
-8\left[ {\hat\partial }_0K_{j[k,l]}+{\hat\partial }_0\Gamma _{j[k}^pK_{l]p}
+{\Gamma }_{j[k}^p{\hat\partial }_0K_{l]p}\right]
{n}^{[0}\bar{m}^{j]}{n}^k\bar{m}^l
\nonumber \\
&&
+4\left[{\hat\partial }_0{R}_{jl}-2K_{(l}^p{\hat\partial }_0K_{j)p}
-2N K_{jp}K_q^pK_l^q\right.
\nonumber \\
&&
\left. +K_{jl}{\hat\partial }_0K+K{\hat\partial }_0K_{jl}\right]
{n}^{[0}\bar{m}^{j]}{n}^{[0}\bar{m}^{l]}
\nonumber \\
&& 
+ 2\{\psi_4 ({l}_i \hat\Delta - m_i\bar{\delta}) N^{i}
+\psi_3 ({n}_i\bar{\delta}-{\bar{m}}_i \hat\Delta) N^{i}\}.
\label{psi4dot}
\end{eqnarray}
where the last term extends the expression in the references, having
been added to take into account the variation of the tetrad terms
$\hat\partial_0\left[{\bf n\bar{m}n\bar{m}}\right]$.  Here
$\hat\Delta=n^\mu\partial_\mu$,
$\bar{\delta}=\bar{m}^\mu\partial_\mu$, and
\begin{eqnarray}
&&\psi _3 = \left[ {R}_{ijkl}+2K_{i[k}K_{l]j}\right]
{l}^i{n}^j{\bar{m}}^k{n}^l \nonumber \\
&&-4\left[ K_{j[k,l]}+{\Gamma }_{j[k}^pK_{l]p}\right]
({l}^{[0}{n}^{j]}{\bar{m}}^k{n}^l
-{n}^{[0}{\bar{m}}^{j]}{l}^k{n}^l)
\label{psi3} \\
&&+2\left[ {R}_{jl}-K_{jp}K_l^p+KK_{jl}\right]
({l}^{[0}{n}^{j]}{\bar{m}}^0{n}^{l}
-{l}^{[0}{n}^{j]}{n}^0{\bar{m}}^{l}), 
\nonumber
\end{eqnarray}
where the background (null and complex) tetrad, (${l}^\mu, {n}^\mu,
{m}^\mu,{\bar{m}}^\mu$) is specified in the subsection below.

The derivatives involved in the above expressions can be computed in
terms of the initial data on the Cauchy hypersurface as in Eq. \
(\protect\ref{derivatives}) in the Appendix.

With the tetrad specified, the foregoing formulae are coordinate
independent. Therefore the only adjustment needed to specify initial
data for the evolution equations we will be to insert the appropriate
background quantities in the above equations.  In particular, taking
$N$ and $N^i$ respectively as the zeroth order Kerr lapse and shift,
$N_{(0)}=\sqrt{\Delta \Sigma/\Omega}$ and
$N_{(0)}^i=[0,0,-2aMr/\Omega]$, allows us to compute
$\partial_t\psi_4$ directly with respect to the background
Boyer-Lindquist time, thus avoiding additional perturbations
introduced if one computes the numerical derivative by finite
differences of $\psi_4$ on two successive slices.

\subsection{The tetrad}\label{tetrad_rotation}

A null and complex `exact' tetrad ({\em i.e.} orthonormal in the
numerical spacetime) must be chosen such that it reduces, in the
linear regime to the choice made in our perturbation treatment of the
final Kerr hole, the Kinnersley tetrad \cite{Teukolsky73}. In
Boyer-Lindquist coordinates the background tetrad vectors are
\begin{subequations}\label{Kin}
\begin{eqnarray}
&&l^\mu_{\rm Kin}=\frac{1}{\Delta}\left[(r^2+a^2),\Delta,0,a\right],
\\
&&n^\mu_{\rm Kin}=\frac{1}{2\Sigma}\left[(r^2+a^2),-\Delta,0,a\right],
\\
&&m^\mu_{\rm Kin}=\frac{1}{\sqrt{2}(r+ia\cos\theta)}
\left[ia\sin\theta,0,1,\frac{i}{sin\theta}\right].
\end{eqnarray}
\end{subequations}
The Kinnersley tetrad is particularly well suited for perturbation
studies because it has the property that $l^\mu$ and $n^\mu$ are
chosen to lie along the (background) principal null directions of the
Weyl tensor (PND) in such a way that one can derive decoupled
perturbation equations.  In terms of the $3+1$ basis of Eqs.\
(\ref{psi4})-(\ref{psi3}), we have
\begin{subequations}\label{Kinhat}
\begin{eqnarray}
&&{l}^\mu=\left[N_{(0)}l^0_{\rm Kin},l^i_{\rm Kin}+N_{(0)}^il^0_{\rm Kin}\right],
\\
&&{n}^\mu=\left[N_{(0)}n^0_{\rm Kin},n^i_{\rm Kin}+N_{(0)}^in^0_{\rm Kin}\right],
\\
&&{m}^\mu=\left[N_{(0)}m^0_{\rm Kin},m^i_{\rm Kin}+N_{(0)}^im^0_{\rm Kin}\right].
\end{eqnarray}
\end{subequations}
To numerically determine an `exact' tetrad we could in principle
search for possible candidates of the two of the PND of the Weyl
tensor.  One could of course, try to pick up some null directions in
our numerical spacetime $g_{\mu\nu}^{num}$ which we know are close to
the PND in Kerr whenever $g_{\mu\nu}^{num}$ is a perturbation of
Kerr. However, this turns out to be a bad choice because the PND do
not behave analytically under analytic perturbations of Kerr. The
reason is that the principal null directions of Kerr are double
principal null directions of the Weyl tensor, which in general will
split under the perturbation.  It turns out that the splitting of
eigenvectors of an endomorphism under a perturbation of order
$\epsilon$ behaves in general as some fractional power of $\epsilon$
(hence non-smoothly).  So, the principal null directions will be too
strongly perturbed.

\def\vone{v_2}
\def\vtwo{v_1}

An alternative and more effective procedure to define an exact tetrad
that has the required property in the linear regime is the following.
(a) We assume the following $3+1$ decomposition of the tetrad
\begin{subequations}\label{PsiK}
\begin{eqnarray}
&&\tilde{l}^\mu=\frac{1}{\sqrt{2}}(u^\mu + r^\mu),\\
&&\tilde{n}^\mu=\frac{1}{\sqrt{2}}(u^\mu - r^\mu),\\
&&\tilde{m}^\mu=\frac{1}{\sqrt{2}}(\theta^\mu+i \varphi^\mu),
\end{eqnarray}
\end{subequations}
where $u^\mu$ is the normalized time-like unit normal to the
hypersurface and $r^\mu=[0,\vone^a],
\theta^\mu=[0,v_3^a], \varphi^\mu=[0,\vtwo^a]$ are orthonormal vectors 
pointing along the numerically defined coordinate directions.
(b) We thus identify the set of null rotations to bring 
\ (\protect\ref{PsiK}) to the form \ (\protect\ref{Kinhat}), in order
to make it consistent with the tetrad assumed in the perturbative 
calculation.
  
Step a is straightforward. Begin with real vectors aligned with the
numerical space's $\varphi$ and radial directions, which in Cartesian
coordinates read
\begin{eqnarray}
&&\vtwo^a=\left[-y,x,0\right],\nonumber\\
&&\vone^a=\left[x,y,z\right],\nonumber\\
&&{v}_3^a=\det(g)^{1/2}g^{ad}\epsilon_{dbc}\vtwo^b\vone^c
\end{eqnarray}
We then redefine these, to achieve ortho-normalization.  It is
important to begin the ortho-normalization procedure with the
azimuthal direction vector $\vtwo^a$ which is not affected by the
frame-dragging effect discussed in Section
\ref{ssec:spatial_coordinates}.  At each step, a Gram-Schmidt
procedure is then used to ensure that the triad remains orthonormal,
so that
\begin{eqnarray}
&&\vtwo^a\rightarrow \frac{\vtwo^a}{\sqrt{\omega_{11}}},
\nonumber\\
&&\vone^a\rightarrow \frac{(\vone^a-\vtwo^a \omega_{12})}{\sqrt{\omega_{22}}},
\nonumber\\
&&{v}_3^a\rightarrow \frac{(v_3^a-\vtwo^a \omega_{13}- \vone^{a} \omega_{23})}
{\sqrt{\omega_{33}}},
\nonumber
\end{eqnarray}
where $\omega_{ij}=(v_i^a v_j^b g_{ab})$.  In the case of Kerr one
finds, in Boyer-Lindquist coordinates,
\begin{subequations}
\begin{eqnarray}
&&u^\mu=\sqrt\frac{\Omega}{\Delta\Sigma}\left[1,0,0,\frac{2aMr}{\Omega}\right],\\
&&r^\mu=[0,\vone^a]=\left[0,\sqrt{\frac{\Delta}{\Sigma}},0,0\right],\\
&&\theta^\mu=[0,v_3^a]=\left[0,0,\frac{1}{\sqrt{\Sigma}},0\right],\\
&&\varphi^\mu=[0,\vtwo^a]=\left[0,0,0,\frac{1}{\sin\theta}\sqrt\frac{\Sigma}
{\Omega}\right],
\end{eqnarray}
\end{subequations}
normalized such that $-u_\mu u^\mu=r_\mu r^\mu=
\theta_\mu \theta^\mu=\varphi_\mu \varphi^\mu=1$ so that the inverse metric
can be expressed as
$g^{\mu\nu}=2(m^{(\mu} \bar{m}^{\nu)}-l^{(\mu} \bar{n}^{\nu)})$

For step b identify a combination of null rotations of type I and II
parameterized by $A$, and a type III (boost) null rotation
parameterized by $F_A$ and $F_B$ which bring the orthonormal tetrad \
(\ref{PsiK}) to the form \ (\ref{Kinhat}) for the unperturbed case.
The transformation
\begin{subequations}\label{transformed-tetrad}
\begin{eqnarray}
l^\mu=&&\frac{F_A}{2}\left[(\sqrt{A^2+1}+1)\ \tilde{l}^\mu
+(\sqrt{A^2+1}-1)\ \tilde{n}^\mu \right.
\nonumber \\
&&
\left.-iA(\tilde{m}^\mu-\tilde{\bar{m}}^\mu)\right],
\\
n^\mu=&&\frac{F_A^{-1}}{2}\left[(\sqrt{A^2+1}-1)\ \tilde{l}^\mu
+(\sqrt{A^2+1}+1)\ \tilde{n}^\mu\right.
\nonumber \\
&&
\left.-iA(\tilde{m}^\mu-\tilde{\bar{m}}^\mu)\right],
\\
m^\mu=&&\frac{F_B}{2}\left[(\sqrt{A^2+1}+1)\ \tilde{m}^\mu
-(\sqrt{A^2+1}-1)\ \tilde{\bar{m}}^\mu\right.
\nonumber \\
&&
\left.+iA(\tilde{l}^\mu+\tilde{n}^\mu)\right].
\end{eqnarray}
\end{subequations}
achieves this with $A= a\sin\theta\sqrt{\Delta/\Omega}$,
$F_A=\sqrt{2\Sigma/\Delta}$ and $F_B=\sqrt{\Sigma}/(r+ia\cos\theta)$,
thereby producing a tetrad consistent with the tetrad assumed in the
perturbative calculation.

In practice we perform the tetrad transformation indirectly ,
implementing its effect on the set of Weyl scalars
$(\psi_0,\ldots,\psi_4)$ as described in the Appendix,
Eq. (\ref{psi-rotation}).

\section{The Teukolsky equation}
 
Perturbations of a rotating Kerr black hole are described by the well
known Teukolsky equation\ \cite{Teukolsky73}, which is derived from
the Newman-Penrose formalism. The Weyl scalar $\psi_4$ that represents
outgoing gravitational radiation satisfies a decoupled wave equation
\begin{eqnarray}\label{generalTeuk}
\Bigg\{&&\left( 
\Delta +4\mu +\overline{\mu }+3\gamma -\overline{\gamma }\right)
\left( D+4\epsilon -\rho \right)\nonumber\\
&&-\left( \overline{\delta }+
3\alpha +\overline{\beta }+4\pi -\overline{\tau }\right)
\left( \delta +4\beta -\tau \right)-3\psi_2^{(0)}\Bigg\}
\psi_4^{(1)}\nonumber\\
&&=0.
\end{eqnarray}
In this generic form the Teukolsky equation is manifestly independent
of the choice of coordinate system used to describe the Kerr
background and its perturbations.  In the foregoing equation the usual
notation for spin coefficients $\alpha, \beta, ...$ was used and
$\hat\Delta=n^\mu\partial_\mu$, $\delta=m^\mu\partial_\mu$, and
$D=l^\mu\partial_\mu$ represent directional derivatives.

For the applications in this paper we consider Boyer--Lindquist
coordinates $(t,r,\theta ,\phi )$ and the Kinnersley tetrad. The
Teukolsky equation then reads
\begin{eqnarray}
&&
\left[{\left(r^2+a^2\right)^2\over\triangle}-a^2\sin^2\theta\right] 
{\partial^2\psi \over \partial t^2}
+{4Mar\over \triangle} {\partial^2 \psi\over\partial t \partial \phi}
 \nonumber\\
&&
+\left[{a^2\over \triangle}-{1 \over\sin^2\theta}\right] 
{\partial^2 \psi \over \partial \phi^2}
-\triangle^2 {\partial \over \partial r} \left( {1 \over \triangle} 
{\partial \psi \over \partial r}\right) 
 \nonumber\\
&&
-{1 \over \sin\theta}{\partial \over \partial\theta} 
\left(\sin\theta {\partial \psi \over \partial \theta}\right)
+4 \left[{M(r^2-a^2)\over \triangle} -r -ia \cos\theta\right] 
{\partial \psi\over \partial t} 
\nonumber\\
&& 
+4 \left[{a(r-M)\over \triangle}+{i \cos\theta\over \sin\theta}\right]
{\partial\psi\over\partial\phi}
+(4\cot^2\theta+2)\psi=0,
\label{Teuk}
\end{eqnarray}
where $\psi =(r-i\, a \cos(\theta))^4\,\psi_4$.

This formulation has several advantages: i) It is a first order gauge
invariant description. ii) It does not rely on any frequency or
multipole decomposition. iii) It can be used to evolve 3+1 dimensional
spacetimes without any assumption about symmetries (to deal with the
final stage of orbiting binary black holes).  iv) The Weyl scalars are
objects defined in the full nonlinear theory and it can be argued that
evolving them with the linear theory provide a reliable description of
the perturbations\cite{Lousto99a}. In addition, the Newman-Penrose
formulation constitutes a simple and elegant framework to organize
higher order perturbation\cite{Campanelli99}.

The numerical integration of the linear Teukolsky equation in the time
domain using Boyer-Lindquist coordinates is done closely following
reference \cite{Krivan97a}. We use the Lax--Wendroff algorithm, using
the standard tortoise coordinate $r^*$,
\begin{eqnarray}
r^*&=&r+\frac{r_+^2+a^2}{r_+-r_-}\ln\left|\frac{r-r_+}{2M}\right|
-\frac{r_-^2+a^2}{r_+-r_-}\ln\left|\frac{r-r_-}{2M}\right|,\nonumber\\
&&r_\pm=M\pm\sqrt{M^2-a^2}
\end{eqnarray}
 which naturally leads to excision of the black hole interior and
constant characteristic wave speed.  We impose static boundary
conditions on the internal boundary (event horizon of the Kerr
background) and radiative boundary conditions on the exterior
boundary. Frame dragging effects are taken care of by the background
Boyer-Lindquist shift.  Thus, this formulation has all the ingredients
to allow for an indefinite stable evolution. In practice it provides
an accurate evolution for the few hundreds of $M$ of relevant signal
generated in the final stages of black hole merger.  Since the Kerr
background has the axial killing vector $\partial_\phi$ we can Fourier
decompose $\psi_4$ into $e^{im\phi_{KS}}$ modes. In particular, for
numerical convenience, we use the `Kerr-Schild' $\phi_{KS}$
\begin{eqnarray}\label{psiKS}
\phi_{KS}&=&\phi+\frac{a}{r_+-r_-}\ln\left|{\frac{r}{r_+}-1}\right|\nonumber\\
&&-\frac{a}{r_+-r_-}\ln\left|{\frac{r}{r_-}-1}\right|.
\end{eqnarray}
This allows us to reduce the dimensionality of the Teukolsky equation
from $3+1$ to $2+1$.  In addition this decomposition into modes can be
applied to the output of the full numerical code with the advantage of
handling $2D$ fields instead of $3D$ ones.  Typical evolutions of the
Teukolsky equation used a grid size of $n_\theta\times
n_{r^*}=40\times1200$, with $-18<r^*/M<78$ for signals of $t\sim100M$,
and we filled in initially with zeroes (or used extrapolations) the
grid-points outside the full numerical domain.  Finally, the
computation of the energy and momenta radiated is performed using the
formulae of Ref.\ \cite{Campanelli99}, Sec.\ III.C.

It worth stressing here that the Teukolsky equation can be written in
any coordinate system. We are using it in the Boyer-Lindquist
coordinates for present convenience, but if full numerical codes
including excision black hole interiors turn out to be more practical
in Kerr-Schild-like coordinates, it may be convenient to evolve
perturbations in a Kerr-Schild slices of the Kerr metric
\cite{Campanelli00a}.

%%%%%%%%%%%%%%%%%%%%%%%%%%%%%%%%%%%%%%%%%%%%%%%%%%%%%%%%%%%%%%%%%%%%%%%%%

\section{Application to a single rotating Kerr hole}\label{Kerr}

We have done extensive testing of our method on various toy models.
In an earlier incarnation, we tried our approach successfully on
axisymmetric head-on collision.  As we have described, we have done a
lot of work generalizing our method to include orbital cases with
angular momentum on the final black hole.  We have checked our
equations explicitly on exact Boyer-Lindquist Kerr data, but in our
real numerical simulations we will not reproduce the Boyer-Lindquist
coordinates exactly and is it useful to get some measure of how
important the coordinate differences are for the radiation.
Additionally, our calculation requires the use of four computer codes,
a code for the numerical simulation ({\sl Cactus} \cite{Cactusweb}), a
specialized code (called {\sl Zorro}) running within the simulation
which calculates the all the quantities needed for producing Cauchy
data (see Appendix), a code (called {\sl TeukCauchy}) which runs after
the simulation defining the background black hole and constructing the
Cauchy data from output of the evolution on various time slices, and
the Teukolsky evolution code ({\sl TeukCode}).  The Kerr case has been
a key source of rigorous tests of all these codes.

We performed a highly nontrivial test of our set up by applying the
entire procedure to Kerr initial data, evolving a single Kerr hole
full numerically, finding a ``background'' black hole in the numerical
data and defining a tetrad, extracting the Cauchy data and continuing
its evolution with the Teukolsky equation. Ideally, the final result
should produce no radiation.  In practice, the computed radiation
energy and waveforms will give us a measure of the error with which we
can determine such quantities.

\begin{figure}[t]
\epsfysize=3.0in \epsfbox{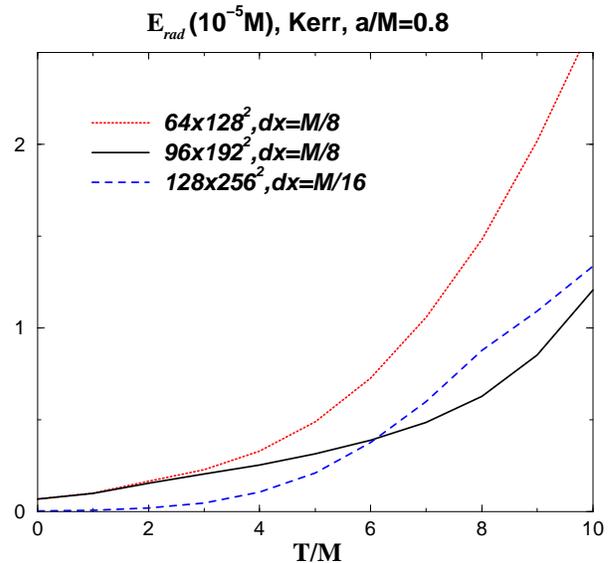}
\caption{The total radiated energy for evolved Kerr hole for different
resolutions and boundary location.}
\label{fig:Kerr_E}
\end{figure}

This is a nontrivial test because the full numerical evolution is
performed with vanishing shift and the singularity avoiding maximal
slicing.  This is in contrast to the Boyer-Lindquist lapse and the
non-vanishing Boyer-Lindquist shift for rotation parameter $a/M=0.8$
in the example shown in Fig.\ \ref{fig:Kerr_E}. In addition the Cauchy
data for $\psi_4$ and $\partial_t\psi_4$ are computed with a tetrad
adapted to the numerical spacetime, which must then be transformed
according to Sec.\ \ref{tetrad_rotation} in order to nearly reproduce
the Kinnersley tetrad in the perturbative limit.  These complications
mean, in particular, that the data passed between our first three
codes is not expected to be approximately vanishing, but must sum to
zero in the end.  In practice our result is subject to both numerical
error and false radiation caused by an inexact identification of the
background coordinates and tetrad, perhaps producing nontrivial Cauchy
data which we then evolve via the Teukolsky equation. The results of
the whole procedure are summarized in Fig.\ \ref{fig:Kerr_E}. The
levels of spurious radiation are around $10^{-5}M$. Results after
relatively short evolution times converge quadratically toward zero
with increasing resolution. The longer evolutions are affected by the
location of a close exterior boundary, and are improved when we move
the boundary outwards by 50\%.  (As discussed above we will use much
more distant outer boundaries for our astrophysical applications.)
All this indicates that the coordinate effects are, so far, smaller
than the numerical effects, which in turn tend to produce radiation
about two orders of magnitude smaller than the radiation we are
interested in. Notably these results are achieved with lower
resolutions and closer boundaries than the typical resolutions of runs
we performed for two black holes starting from the ISCO configuration
used in Ref.\ \cite{Baker:2001nu}.

For the sake of completeness we mention two further tests which we
performed for two black hole initial data: i) The mass scaling of the
whole procedure. Since Einstein equations scale with the total ADM
mass, we made a full numerical run with initial mass equal 2 and
compare the scaling of the Cauchy data, post-processing and final
waveforms with the mass equal 1 case. This proved to be a very useful
test for the corresponding set of four codes we used to compute each
of the above stages. ii) Reducing the initial separation of the holes
from that of the ISCO to one quarter of it we reach the close limit
regime and can compare with the results with the known analytic
expressions \cite{Khanna:2000dg} and scaling with the separation as
well as angular dependence of $\psi_4$ and $\partial_t\psi_4$.

\section{Discussion}
%We identify the numerical ``maximal'' slicing $(K=0)$, with the
%Boyer-Lindquist time, and tested this was a good approximation
%outside the final common event horizon of the binary system.
%We also identify the $\theta$ coordinate with the numerical one.
%To correct for the ``grid stretching'' we derive the radial
%coordinate from the invariant
%${\cal I}_{Kerr}=3M^3/(r-ia\cos\theta)^6$, and the $\varphi$
%coordinate form the condition $g_{r\varphi}=0$ to account for
%frame dragging effects.

The Lazarus approach to binary black holes combines three treatments,
each adapted to one of three stages of the dynamics, the far-limit,
non-linear-interaction, and close-limit (one black hole) regimes.  In
this paper we have provided a detailed explanation of how numerical
simulation and Teukolsky equation perturbative dynamics can be
interfaced to provide a complete description of gravitational
radiation arising from the post-orbital binary black hole dynamics.

This technology makes it possible, for the first time, to apply
numerical relativity, to the non-linear dynamical interaction of these
systems. In our approach to this unknown regime, we have identified
several parameter sequences which make a connection to better-studied
cases.  An ``L-sequence'', allows us to increase the separation from
close-limit regime to ISCO, an ``$\alpha$-sequence'' which allows us
to connect boosted head-on collisions studied in 2D to the ISCO with a
fixed magnitude of each black hole's momentum, and a ``P-sequence''
through which we connect to head-on collisions of resting black holes
by varying the magnitude of the momentum to the ISCO value
\cite{Baker:2001nu}. These important studies will give us some
understanding of how the dynamics from ISCO configurations relate to
and differ from the simpler problems treated so far by numerical
simulations and close-limit studies.

After establishing such a basis for understanding the near ISCO regime
dynamics we can reach out further, and seek to firmly establish the
relation of these ISCO black hole configuration to astrophysics.  As
we have discussed in detail for the numerical simulation/close-limit
interface, the dual approach to dynamics in overlapping validity
regions provides a vital consistency check on the reliability of the
results.  A key goal, which we can now begin to approach, is to
provide the same sort of consistency studies to the
far-limit/numerical simulation interface and to thereby establish a
firm astrophysical foundation for expensive, and difficult, numerical
work.  Again, a handful of initial data sequences are appropriate for
beginning to evaluate the connection of numerical/close-limit results
to astrophysical problems.  Within the effective-potential method
which we have taken advantage of in order to define of ISCO data are a
natural ``PI-sequence'' of pre-ISCO stable circular orbits.  Moving up
this sequence toward more separated black holes asymptotically
eliminates the features of these data which may be less astrophysical.
Similarly, it is possible to define curves through the parameter space
of our initial data family which approaches the trajectories defined
by the Buonanno-Damour extension of the post-Newtonian method.  The
application of more advanced numerical techniques
\cite{Alcubierre01a}, which we are presently undertaking, should make
it possible to begin generating waveforms from farther up these
sequences.  Still, though, the effective potential method initial data
sequence is an imperfect stand-in for robust interface with the
post-Newtonian method, which we expect to be ultimately required.
Since a primary concern about the astrophysical relevance of
numerical/close-limit results is artificial radiation content in the
initial data; another useful line of research is comparison studies of
waveforms alternative initial data sets which would be equivalent in
their astrophysical interpretation.  These will provide a measure of
significance of this interpretive indeterminacy to the gravitational
radiation.  Promising work with evolutions from Kerr-Schild-like
initial data, for which an alternative instantiation of the Lazarus
approach is under development \cite{PSU-Laz}, should enable an example
of such comparative work.

Another area of study which can now be pursued is to develop some
preliminary indications of the effect spin has on the waveforms
generated in the post-inspiral dynamics.  The effective potential
approach provides a description \cite{Pfeiffer:2000um} of the effect
small amounts of individual black hole spin have of the ISCO initial
data.  We are applying our approach at first instance to cases of spin
parallel and antiparallel to the orbital angular momentum.

Eventually numerical simulations will run routinely for hundreds or
thousands of $M$, having begun from established astrophysical
data. But the possibility for observation is beginning almost
immediately, and until now we have not met the needs of observers who
express that any additional information about the the final stage of
binary black holes may be extremely important \cite{Damour00a}. Our
efforts have shown that a crucial requirement for producing results
relevant to observers is to adapt numerical evolutions to
astrophysical problems.  Numerical relativity is ready, now, to begin
answering questions about binary black holes in the near ISCO and
pre-ISCO regime.

%%%%%%%%%%%%%%%%%%%%%%%%%%%%%%%%%%%%%%%%%%%%%%%%%%%%%%%%%%%%%%%%%%%%%%%%%

\begin{acknowledgments}
We wish to specially thank B. Br\"ugmann and R. Takahashi for many
contributions to this work. We also acknowledge P. Laguna, R. Price,
E. Seidel, and B. Schutz for many helpful discussions.  M.C. was also
supported by the IHRP program of the European Union (Marie-Curie
Fellowship HPMF-CT-1999-00334).  All our numerical computations have
been performed on AEI and NCSA Origin 2000s.

We would also like to thank David Fiske and Breno Imbiriba for
pointing out errors in the published paper, and Bernard Kelly for
correcting these and other inconsistencies in this revision of the
text.
\end{acknowledgments}

\appendix
\section{A practical construction of the Cauchy data }

Here we describe a procedure for calculating $\psi_4$ which allows us to
cleanly separate the specification of the background black hole from
the numerical simulation, as is practical for studying variations of the
background metric and background coordinates.
Since we have not yet determined the background black hole at the time of
evolution we must compute a larger set of quantities which can then be 
transformed to the desired result after the background is specified.
The steps are:

(a) 
Compute the numerical $3+1$ tetrad components as in 
Eq.\ (\ref{PsiK})

(b) 
With this tetrad, using the Cauchy data on a numerical time slice 
directly compute all five Weyl scalars 
$\psi_{0}\ldots\psi_{4}$ and their time 
variations $\hat\partial_0\psi_{0}\ldots\hat\partial_0\psi_{4}$ 
as follows: 
\begin{eqnarray}
&&
\psi_4=\R_{ijkl}n^i\bar{m}^jn^k\bar{m}^l\nonumber \\&&
+2\R_{0jkl}(n^0\bar{m}^jn^k\bar{m}^l-\bar{m}^0n^jn^k\bar{m}^l)\nonumber \\&&
+\R_{0j0l}(n^0\bar{m}^jn^0\bar{m}^l+\bar{m}^0n^j\bar{m}^0n^l
-2n^0\bar{m}^j\bar{m}^0n^l)
\nonumber \\\\
&&
\psi_3=\R_{ijkl}l^in^j\bar{m}^kn^l\nonumber \\&&
+\R_{0jkl}(l^0n^j\bar{m}^kn^l-n^0\bar{m}^jl^kn^l-n^0l^j\bar{m}^kn^l+
         \bar{m}^0n^jl^kn^l)\nonumber \\&&
+\R_{0j0l}(l^0n^j\bar{m}^0n^l-l^0n^jn^0\bar{m}^l-n^0l^j\bar{m}^0n^l+
         n^0l^jn^0\bar{m}^l)
\nonumber \\\\
&&
\psi_2=\R_{ijkl}l^im^j\bar{m}^kn^l\nonumber \\ 
&&
+\R_{0jkl}(l^0m^j\bar{m}^kn^l-n^0\bar{m}^jl^km^l
-m^0l^j\bar{m}^kn^l+\bar{m}^0n^jl^km^l)\nonumber \\
&&
+\R_{0j0l}(l^0m^j\bar{m}^0n^l-l^0m^jn^0\bar{m}^l
-m^0l^j\bar{m}^0\bar{m}^l+n^0l^jm^0\bar{m}^l)
\nonumber \\\\
&&
\psi_1=\R_{ijkl}n^il^jm^kl^l\nonumber \\&&
+\R_{0jkl}(n^0l^jm^kl^l-l^0m^jn^kl^l-l^0n^jm^kl^l+
         m^0l^jn^kl^l)\nonumber \\&&
+\R_{0j0l}(n^0l^jm^0l^l-n^0l^jl^0m^l-l^0n^jm^0l^l+l^0n^jl^0m^l)
\nonumber \\\\
&&
\psi_0=\R_{ijkl}l^im^jl^km^l\nonumber \\&&
+2\R_{0jkl}(l^0m^jl^km^l-m^0l^jl^km^l)\nonumber \\&&
+\R_{0j0l}(l^0m^jl^0m^l+m^0l^jm^0l^l-2l^0m^jm^0l^l)
\nonumber \\
\end{eqnarray}
where $\R_{abcd}$ is the four-dimensional Riemann tensor:
\begin{eqnarray}
&&
\R_{ijkl}={R}_{ijkl}+2K_{i[k}K_{l]j} 
\nonumber \\
&&
\R_{0jkl}=-2\left[K_{j[k,l]}+{\Gamma }_{j[k}^pK_{l]p}\right]
\nonumber \\
&&
\R_{0j0l}={R}_{jl}-K_{jp}K_l^p+KK_{jl}
\end{eqnarray}

We compute the time variations only from data on the slice:
\begin{eqnarray}
&&
\hat\partial_0\psi_4=\hat\partial_0\R_{ijkl}n^i\bar{m}^jn^k\bar{m}^l
\nonumber \\
&&
+2\hat\partial_0\R_{0jkl}(n^0\bar{m}^jn^k\bar{m}^l-\bar{m}^0n^jn^k\bar{m}^l)
\nonumber \\
&&
+\hat\partial_0\R_{0j0l}(n^0\bar{m}^jn^0\bar{m}^l+\bar{m}^0n^j\bar{m}^0n^l
-2n^0\bar{m}^j\bar{m}^0n^l)
\nonumber \\\\
&&
\hat\partial_0\psi_3=\hat\partial_0\R_{ijkl}l^in^j\bar{m}^kn^l
\nonumber\\
&&
+\hat\partial_0\R_{0jkl}(l^0n^j\bar{m}^kn^l-n^0\bar{m}^jl^kn^l
-n^0l^j\bar{m}^kn^l+\bar{m}^0n^jl^kn^l)
\nonumber \\&&
+\hat\partial_0\R_{0j0l}(l^0n^j\bar{m}^0n^l-l^0n^jn^0\bar{m}^l
-n^0l^j\bar{m}^0n^l+n^0l^jn^0\bar{m}^l)
\nonumber \\\\
&&
\hat\partial_0\psi_2=\hat\partial_0\R_{ijkl}l^im^j\bar{m}^kn^l
\nonumber \\
&&
+\hat\partial_0\R_{0jkl}(l^0m^j\bar{m}^kn^l-n^0\bar{m}^jl^km^l
-m^0l^j\bar{m}^kn^l+\bar{m}^0n^jl^km^l)\nonumber \\&&
+\hat\partial_0\R_{0j0l}
(l^0m^j\bar{m}^0n^l-l^0m^jn^0\bar{m}^l
-m^0l^j\bar{m}^0\bar{m}^l+n^0l^jm^0\bar{m}^l)
\nonumber \\\\
&&
\hat\partial_0\psi_1=\hat\partial_0\R_{ijkl}n^il^jm^kl^l
\nonumber\\
&&
+\hat\partial_0\R_{0jkl}(n^0l^jm^kl^l-l^0m^jn^kl^l-l^0n^jm^kl^l+
m^0l^jn^kl^l)
\nonumber \\
&&
+\hat\partial_0\R_{0j0l}(n^0l^jm^0l^l-n^0l^jl^0m^l-l^0n^jm^0l^l+l^0n^jl^0m^l)
\nonumber \\\\
&&
\hat\partial_0\psi_0=\hat\partial_0\R_{ijkl}l^im^jl^km^l
\nonumber \\
&&
+2\hat\partial_0\R_{0jkl}(l^0m^jl^km^l-m^0l^jl^km^l)
\nonumber \\
&&
+\hat\partial_0\R_{0j0l}(l^0m^jl^0m^l+m^0l^jm^0l^l-2l^0m^jm^0l^l).
\nonumber \\
\end{eqnarray}
The derivatives involved in the above expressions can be computed 
in terms of the data on the Cauchy hypersurface using Einstein's
equations,
\begin{eqnarray}
&&
\hat\partial_0\R_{ijkl} =-4N \big\{ K_{i[k}{R}_{l]j}-K_{j[k}{R}_{l]i}  
\nonumber\\
&& 
-\frac 12{R}\left(K_{i[k}g_{l]j}-K_{j[k}g_{l]i}\right) \big\}
\nonumber\\
&&
+2g_{i[k}\hat\partial_0{R}_{l]j}-2g_{j[k}
\hat\partial_0{R}_{l]i}-g_{i[k}g_{l]j}{\partial }_0{R}
\nonumber\\
&&
+2K_{i[k}{\partial }_0K_{l]j}-2K_{j[k}{\partial }_0K_{l]i},
\nonumber\\\\
&&
\hat\partial_0\R_{0jkl} =   
{\partial }_0K_{j[k,l]}+{\hat\partial }_0\Gamma _{j[k}^pK_{l]p}
+{\Gamma }_{j[k}^p{\hat\partial }_0K_{l]p},
\nonumber\\\\
&&
\hat\partial_0\R_{0j0l} = 
\left[{\partial }_0{R}_{jl}-2K_{(l}^p{\partial }_0K_{j)p}\right.
\nonumber \\
&&
\left. -2N K_{jp}K_q^pK_l^q+K_{jl}{\partial }_0K+K{\partial}_0K_{jl}
\right]
\end{eqnarray}
where
\begin{eqnarray}
&&
\hat\partial_0K=N K_{pq}K^{pq}-{\nabla }^2N, 
\nonumber\\\\&& 
\hat{\partial }_0K_{ij}=N\left[ R
_{ij}+KK_{ij}-2K_{ip}K^p{}_j-N^{-1}\nabla _i\nabla
_jN\right].
\nonumber\\\\&&
\hat\partial_0{R}=2K^{pq}{\partial }_0K_{pq}
+4NK_{pq}K_s^pK^{sq}-2K{\partial }_0K
\nonumber\\\\&&
\widehat{\partial }_0R_{ij}=\nabla _k(\widehat{\partial }_0
\Gamma _{ij}^k)-\nabla _j(\widehat{\partial }_0\Gamma _{ik}^k),
\nonumber\\\\&&
\widehat{\partial }_0\Gamma _{ij}^k=-2\nabla _{(i}(NK_{j)}{}^k)+
\nabla ^k(NK_{ij}).  \label{derivatives}
\end{eqnarray}

These time variations may be precisely the background time derivatives we
want if have already specified the background spacetime and can set
$N=N_{(0)}$, the background lapse.  On the other hand when we want to
determine the background independently of the numerical simulation code we 
can produce the required information for later processing from all five 
quantities with the choice $N=1$.

(c)
Next we define the background coordinate system as described in
Section \ref{coordinates}.  This allows us now to refer to quantities
defined in the background coordinates.  If we have used $N=1$ in
constructing the time variations we must next translate the time
variations calculated above to genuine background time derivatives
with a set of corrections for the effect of the background lapse and
shift. For the lapse,
\def\f{\alpha_{,\hat r}\,}
\def\g{\alpha_{,\hat \theta}\,}
\def\Br{\beta^{\hat \varphi}_{,\hat r}\,}
\def\Bt{\beta^{\hat \varphi}_{,\hat \theta}\,}
\begin{subequations}\label{lapsecorrection}
\begin{eqnarray}
\hat\partial_0\psi_0&\rightarrow&\alpha\hat\partial_0\psi_0+2\,\f\psi_0+2\,\g\psi_1\\
\hat\partial_0\psi_1&\rightarrow&\alpha\hat\partial_0\psi_1+\f\psi_1+\frac{1}{2}\,\g(\psi_0+3\psi_2)\\
\hat\partial_0\psi_2&\rightarrow&\alpha\hat\partial_0\psi_2+\g(\psi_1+\psi_3)\\
\hat\partial_0\psi_3&\rightarrow&\alpha\hat\partial_0\psi_3-\f\psi_3+\frac{1}{2}\,\g(\psi_4+3\psi_2)\\
\hat\partial_0\psi_4&\rightarrow&\alpha\hat\partial_0\psi_4-2\,\f\psi_4+2\,\g\psi_3
\end{eqnarray} 
\end{subequations}
where $\f$ and $\g$ are related to the lapse for the background Kerr
metric,
\begin{equation}
\begin{array}{rcccl}
\f&=&\frac{1}{\sqrt{g_{rr}}}N_{(0),r}&=&\sqrt{\frac{\Delta}{\Sigma}}N_{(0),r}\cr
\g&=&\frac{1}{\sqrt{g_{\theta\theta}}}N_{(0),\theta}&=&\sqrt{\frac{1}{\Sigma}}
N_{(0),\theta}.
\end{array}
\end{equation}

The shift corrections are
\begin{subequations}\label{shiftcorrection}
\begin{eqnarray}
\hat\partial_0\psi_0&\rightarrow\hat\partial_0\psi_0&+i\,\Bt\psi_0+i\,\Br\psi_1+N^k\partial_k\psi_0\\
\hat\partial_0\psi_1&\rightarrow\hat\partial_0\psi_1&+\frac{i}{2}\,\Br(\psi_0+\psi_2)
                                 -\frac{i}{2}\Bt(\psi_1-\bar\psi_1)\nonumber\\
                               &&+\frac{i}{2}\Br(\psi_2-\bar\psi_2)+N^k\partial_k\psi_1\\
\hat\partial_0\psi_2&\rightarrow\hat\partial_0\psi_2&+\frac{i}{2}\,\Br(\psi_1+\psi_3)+N^k\partial_k\psi_2\\
\hat\partial_0\psi_3&\rightarrow\hat\partial_0\psi_3&+\frac{i}{2}\,\Br(\psi_4+\psi_2)
                                +\frac{i}{2}\Bt(\psi_3-\bar\psi_3)\nonumber\\
                              &&+\frac{i}{2}\Br(\psi_2-\bar\psi_2)+N^k\partial_k\psi_3\\
\hat\partial_0\psi_4&\rightarrow\hat\partial_0\psi_4&-i\,\Bt\psi_4+i\,\Bt\psi_3+N^k\partial_k\psi_4
\end{eqnarray} 
\end{subequations}
where $\Br$ and $\Bt$ are related to the shift for the background Kerr
metric,
\begin{equation}
\begin{array}{rcccl}
\Br&=&\sqrt{\frac{g_{\varphi\varphi}}{g_{rr}}}{N^\varphi}_{(0),r}
&=&\sqrt{\Delta \Omega}\frac{\sin{\theta}}{\Sigma}{N^\varphi}_{(0),r}\cr
\Bt&=&\sqrt{\frac{g_{\varphi\varphi}}{g_{\theta\theta}}}
{N^\varphi}_{(0),\theta}&=&\sqrt{\Omega}\frac{\sin{\theta}}{\Sigma}
{N^\varphi}_{(0),\theta}.
\end{array}
\end{equation}

(d)

The Weyl scalars corresponding to the transformed tetrad defined in
Eq.\ (\ref{transformed-tetrad}) , $\tilde{\psi}_4$ and
$\partial_t\tilde{\psi}_4$ can then be respectively expressed as a
linear combination of the five numerical-tetrad Weyl scalars
$\psi_{4}\ldots\psi_{0}$ and the
$\hat\partial_0\psi_{4}\ldots\hat\partial_0\psi_{0}$, as given in
Eqs.(\ref{shiftcorrection}), with coefficients depending on the
background coordinates $r$ and $\theta$, and on $M$ and $a$:
\begin{eqnarray}
&&
\tilde{\psi_4}=\frac{1}{4F_A^2F_B^2}\big[(\sqrt{A^2+1}-1)^2\psi_0
\nonumber \\
&&
+4iA(\sqrt{A^2+1}-1)\psi_1-6A^2\psi_2
\nonumber \\
&&
-4iA(\sqrt{A^2+1}+1)\psi_3+(\sqrt{A^2+1}+1)^2\psi_4\big]
\label{psi-rotation}\\
&&
\partial_t\tilde{\psi_4}=\frac{1}{4F_A^2F_B^2}\big[(\sqrt{A^2+1}-1)^2\hat\partial_0\psi_0
\nonumber \\
&&
+4iA(\sqrt{A^2+1}-1)\hat\partial_0\psi_1-6A^2\hat\partial_0\psi_2
\nonumber \\
&&
-4iA(\sqrt{A^2+1}+1)\hat\partial_0\psi_3+(\sqrt{A^2+1}+1)^2\hat\partial_0\psi_4\big]
\nonumber
\end{eqnarray}

(e)

We use $e^{i m\phi_{KS}}$ decomposition which is affected by the
$\varphi$ transformation given in Eqs. (\ref{phi-transformation}) and
(\ref{psiKS}).  This transformation is implemented at the end of the
calculation by $\tilde\psi_4\rightarrow e^{i m
\varphi{}_{offset}}\tilde\psi_4$ and likewise for
$\partial_t\tilde\psi_4$. Note that the calculation of the last shift
correction term in Eqs. (\ref{shiftcorrection}) can be also
conveniently carried over after step (e) rather than in step (d).

\bibliographystyle{prsty}
%\bibliography{bibtex/references}
%\thebibliography{PRD}
\bibliography{PRD}

\end{document}